\documentclass[aps,prl,twocolumn]{revtex4}

\usepackage{amssymb,amsmath,graphicx,graphics,color,epstopdf}

\newcommand{\ket}[1]{|#1\rangle}
\newcommand{\bra}[1]{\langle#1|}

\newcommand{\braket}[2]{\langle#1|#2\rangle}

\newcommand{\schr}{Schr\"odinger }

\begin{document}

\title{Complex Berry phase instability in $\mathcal{PT}$-symmetric coupled waveguides}
\author{Rosie Hayward}
\author{Fabio Biancalana}
\affiliation{School of Engineering and Physical Sciences, Heriot-Watt University, EH14 4AS Edinburgh, UK}

\begin{abstract}
We show that the analogue of the geometric phase for non-Hermitian coupled waveguides with $\mathcal{PT}$-symmetry and at least one periodically varying parameter can be purely imaginary, and will consequently result in the manifestation of an instability in the system. 
The instability peaks seen in the spectrum of the system's eigenstates after evolution along the waveguides can be directly mapped to the spectrum of the derivative of the geometric function. 
The instabilities are magnified as the exceptional point of the system is approached, and non-adiabatic effects begin to appear. As the system cannot evolve adiabatically in the vicinity of the exceptional point, $\mathcal{PT}$-symmetry will be observed breaking earlier than theoretically predicted. 
 
\end{abstract}

\maketitle

\section{Introduction}

Since the concept of a geometric phase in quantum mechanics was first described by Berry in 1984 \cite{Berry1}, it has been extended such that it can be applied in many areas of physics, including to systems which are non-adiabatic and non-Hermitian \cite{GandW, AandA}. As such, the effect it has on such systems is of considerable interest, in particular to the optics community. The analogue of the Berry phase in a non-Hermitian system is easily derived via the traditional approach \cite{Berry2}, through use of Floquet theorem \cite{Floquet1, Floquet2}, or by using an evolution operator method \cite{Dattoli}. It is known that this geometric `phase' is not necessarily a real function for non-Hermitian systems, and as such it is possible for eigenstates to gain a real exponential multiplier, rather than simply a phase, after cyclic adiabatic evolution \cite{Dattoli, GandW}, leading to an exponential growth of the amplitude, i.e. an instability. 

By replacing the requirement that a Hamiltonian be Hermitian with the weaker condition of parity-time ($\mathcal{PT}$) symmetry, a type of space-time reflection symmetry, it is possible for it to have real eigenvalues as long as $\mathcal{PT}$-symmetry remains unbroken \cite{Bender1, Bender2, BenderBerry}. In the language of quantum mechanics, when $\mathcal{PT}$-symmetry is unbroken, the parity-time operator will commute with the Hamiltonian, and share its instantaneous eigenstates; it breaks when the eigenstates of the Hamiltonian are no longer eigenstates of the $\mathcal{PT}$ operator \cite{Bender1}. Such non-Hermitian systems are easily realised in the field of optics, for example as coupled waveguides with balanced gain and loss \cite{Christo1, Christo2}. The $\mathcal{PT}$ symmetry breaking point of such a system corresponds to an exceptional point in parameter space, where the eigenvalues and eigenstates of the system coalesce  \cite{Rotter1, Christo1}. If the parameters cross this point, the eigenvalues of the system can become purely imaginary, and exponential gain will be observed in the waveguides. 

Exceptional points have gained a lot of interest in recent years due to their various possible applications \cite{Rotter1, Christo3}, and as such there have been interesting results concerning the geometric phase and the onset of instability in the non-Hermitian systems. Theoretically, the adiabatic encircling of an exceptional point can lead to a state-flip or the accumulation of a geometric phase in a two-level system \cite{Rotter1}. However, in practice adiabaticity breaks down, as even with slow evolution the gain inherently present in such a system will magnify non-adiabatic effects which are usually neglected \cite{Uzdin}. This is indeed what is observed in the work that follows. Furthermore, optical non-Hermitian systems near the exceptional point are known to be unstable with respect to infinitesimally small changes in the system's parameters \cite{Zyablovsky}. 

In this work, we show that a non-Hermitian, $\mathcal{PT}$-symmetric coupled waveguide system with balanced gain and loss and a periodically varying coupling will have a purely imaginary geometric phase below the exceptional point. On approaching the exceptional point, adiabaticity will break down even for a slowly varying coupling between the waveguides, and it is consequently possible for the broken $\mathcal{PT}$-symmetry phase to occur earlier in parameter space than expected, due to the non-adiabatic `drift' of the energy eigenstates. It is, nevertheless, still possible to directly observe the contribution of the non-Hermitian geometric `phase' after cyclic evolution in the form of an instability appearing in Fourier space, due to the fact that the instantaneous eigenvectors will gain an exponential change which no longer simply amounts to a phase factor.

\section{The Geometric Phase for Non-Hermitian Systems}

Complex Hamiltonians, such as those containing gain and loss terms, will not in general be Hermitian, and will have a set of eigenvectors $\ket{\psi_n(z)}$ and adjoint eigenvectors $\ket{\phi_n(z)}$, such that:
\begin{equation}\label{eq:h1}
i\frac{d}{dz}\ket{\psi_n(z)}=\hat{H}(z)\ket{\psi_n(z)},
\end{equation}
and
\begin{equation}\label{eq:h2}
i\frac{d}{dz}\ket{\phi_n(z)}=\hat{H}^{\dagger}(z)\ket{\phi_n(z)},
\end{equation}
which satisfy the following stationary equations:
\begin{equation}
\hat{H}(z)\ket{\psi_n(z)}=\lambda_n\ket{\psi_n(z)},
\end{equation}
and
\begin{equation}
\hat{H}^{\dagger}(z)\ket{\phi_n(z)}=\lambda_n^*\ket{\phi_n(z)}.
\end{equation}

The eigenvectors and adjoint-eigenvectors are bi-orthogonal and complete, such that $\braket{\phi_m(z)}{\psi_n(z)}=0$ for $m\neq n$, and the condition \cite{Brody},
\begin{equation}
\sum_n\frac{\ket{\psi_n(z)}\bra{\phi_n(z)}}{\braket{\phi_n(z)}{\psi_n(z)}}=\hat{\mathbb{I}},
\end{equation}
is satisfied, where $\hat{\mathbb{I}}$ is the identity operator.

Often in the literature, the normalisation convention $\braket{\phi_n(z)}{\psi_n(z)}=1$ is enforced. However, normalising the eigenvectors in this way when dealing with a system with exceptional or diabolical points can be problematic \cite{Rotter1, Keck}. Although the inner products of states in quantum mechanical Hermitian systems relate to probabilistic interpretations of measurement outcomes, the same interpretation cannot be applied to complex Hamiltonians \cite{Brody}. As our system is optical, rather than quantum mechanical, and as one does not need to normalise the eigenvectors and adjoint eigenvectors to calculate the geometric phase, normalising the states is unnecessary, and hence the convention is not enforced in this work.

For a general solution to the \schr equation, $\ket{\Psi(z)}=\sum_n c_n(z) \ket{\psi_n(z)}e^{-i\int_0^z \lambda_n(z')dz'}$, we can solve to find the $z$-evolution of the coefficients:
\begin{multline}
\bigg(\hat{H}(z)-i\frac{d}{dz}\bigg)\ket{\Psi(z)}=\\i\sum_n\big( \dot{c}_n \ket{\psi_n}+ c_n \frac{d}{dz}\ket{\psi_n}\big)e^{-i\int_0^z \lambda_n(z')dz'}=0.
\end{multline}
As we have a bi-orthogonal system, we take the inner product of the above with a particular adjoint-eigenvector, $\ket{\phi_n}e^{-i\int_0^z \lambda^*_n(z')dz'}$, rather than an eigenvector, and find the condition:
\begin{multline}\label{eq:coef}
i \dot{c}_n \braket{\phi_n}{\psi_n}=-c_n\bra{\phi_n}i  \frac{d}{dz}\ket{\psi_n}\\-c_m\bra{\phi_n}i\frac{d}{dz}\ket{\psi_m}e^{-i\int_0^z [\lambda_n(z')-\lambda_m(z')]dz'}.
\end{multline}

From here, if we assume the adiabatic approximation is upheld, we can neglect the $c_m$ term above, since the oscillating phase will average out to zero in case of slow evolution, to get
\begin{equation}
 \dot{c}_n= c_n i \, \frac{\bra{\phi_n}i  \frac{d}{dz}\ket{\psi_n}}{\braket{\phi_n}{\psi_n}},
\end{equation}
which implies we can write the expansion coefficient $c_n$ as an exponential phase factor: $e^{i \gamma_b(z)}$, and deduce that the geometric phase for our bi-orthogonal system is given by \cite{Berry2},
\begin{equation}\label{eq:berry}
\gamma_b(z)=\int_0^z \frac{\bra{\phi_n}i \frac{d}{dz'}\ket{\psi_n}}{\braket{\phi_n}{\psi_n}}dz'.
\end{equation}

Unlike the Berry phase, it is possible for the above to become purely imaginary, leading to eigenstates gaining a real, exponential multiplier during evolution. We must, of course, have a parameter space with a nontrivial topology to gain a result of interest. 

Of course we must also consider the secondary term in (\ref{eq:coef}), proportional to the complex exponential, which can only be neglected for adiabatic evolution. In analogy with the standard procedure for Hermitian systems, we can assume the adiabatic approximation takes the form ($\hbar=1$) \cite{GandW, Nesterov}:
\begin{equation}\label{eq:adiabatic}
\bigg| \bra{\phi_n}i \frac{d}{dz}\ket{\psi_m}\bigg|\ll|\lambda_m(z)-\lambda_n(z)|,
\end{equation}
which can be rewritten in the following way:
\begin{equation}\label{eq:adiabatic}
\bigg| \frac{\bra{\phi_n}i \frac{d \hat{H}(z)}{dz}\ket{\psi_m}}{\lambda_m(z)-\lambda_n(z)}\bigg|\ll|\lambda_m(z)-\lambda_n(z)|,
\end{equation}
i.e. we require that the phase is evolving rapidly along $z$, and consquently the eigenvalue separation to be large, with respect to the change in $z$ of the Hamiltonian.

The above condition should not be considered to always hold true in non-Hermitian systems, as they are known to exhibit quasiadiabatic dynamical effects near exceptional points \cite{Rotter2}. Consequently, even when the inequality (\ref{eq:adiabatic}) is upheld, we can expect that the instantaneous eigenstates may not evolve such that the final state of the system remains an instantaneous eigenstate, varying only by a complex phase factor.

\section{PT-Symmetric Coupled Waveguides with a Periodic Coupling}

\subsection{Geometric phase}
Consider $\mathcal{PT}$-symmetric linearly coupled waveguides with balanced gain and loss, as seen in \cite{Christo1}, adapted so that the coupling between the waveguides varies periodically along with propagation direction, $z$:
\begin{subequations}\label{eq:coup}
\begin{align}
i\frac{d \psi_1}{dz}+ \kappa(z) \, \psi_2-i \, \gamma \, \psi_1=0; \\
i\frac{d \psi_2}{dz}+ \kappa(z) \, \psi_1+i \, \gamma \, \psi_2=0; 
\end{align}
\end{subequations}
where $\psi_1$ and $\psi_2$ represent the modal field amplitude in each channel, respectively, $k(z)$ is the periodic coupling between the waveguides, and $\gamma$ is a scaled gain (loss) coefficient. This problem can be recast in the style of a quantum mechanical two-level system with Hamiltonian
\begin{equation}\label{eq:ham}
\hat{H}=\begin{pmatrix} i \gamma && -\kappa(z) \\ -\kappa(z) && -i \gamma \end{pmatrix},
\end{equation}
with instantaneous eigenvectors of the form $\ket{\psi_n}=\frac{1}{\sqrt{2}}\begin{pmatrix} \psi_1 \\ \psi_2 \end{pmatrix}$. It is easy to analytically solve the above. Making the substitution $\gamma/k(z)=\sin{(\alpha(z))}$, the eigenvectors can be concisely written as,
\begin{equation}
\ket{\psi_1}=\frac{1}{\sqrt{2}}\begin{pmatrix} e^{-i \frac{\alpha}{2}} \\ e^{i \frac{\alpha}{2}} \end{pmatrix}, \, \ket{\psi_2}=\frac{1}{\sqrt{2}}\begin{pmatrix} i \, e^{i \frac{\alpha}{2}} \\ -i \, e^{-i \frac{\alpha}{2}} \end{pmatrix},
\end{equation}
with corresponding eigenvalues,
\begin{equation}
\lambda_{1,2}=\mp \sqrt{\kappa^2(z)-\gamma^2}.
\end{equation}

The adjoint-eigenvectors can also be found by solving for the eigensystem of $\hat{H}^{\dagger}$:
\begin{equation}
\ket{\phi_1}=\frac{1}{\sqrt{2}}\begin{pmatrix} e^{i \frac{\alpha}{2}} \\ e^{-i \frac{\alpha}{2}} \end{pmatrix}, \, \ket{\phi_2}=\frac{1}{\sqrt{2}}\begin{pmatrix} i \, e^{-i \frac{\alpha}{2}} \\ -i \, e^{i \frac{\alpha}{2}} \end{pmatrix},
\end{equation}
As the eigenvalues are real below the $\mathcal{PT}$ breaking point, the eigenvalues of the adjoint system are identical and real: $\lambda_{1,2}^*=\lambda_{1,2}$.

It is easy to confirm that this system is $\mathcal{PT}$-symmetric, as the Hamiltonian (\ref{eq:ham}) commutes with the $\mathcal{PT}$ operator, as in \cite{Christo1}. In this case, the application of the $\mathcal{PT}$ operator corresponds to the transformation $\sigma_x \hat{H}^*\sigma_x$, where $\sigma_x$ is the first Pauli matrix. This Hamiltonian can have an entirely real spectrum below the point where the system transitions to the broken $\mathcal{PT}$-symmetric phase. This point is crossed when the eigenvalues and eigenvectors coalesce, which in our case is when $\kappa(z)=\pm\gamma$ and $\lambda_{1,2}=0$. This is known as the {\em exceptional point} of the system \cite{Rotter1}.

For (assumed) adiabatic evolution of the eigenstates, we can find $\gamma_b$ using (\ref{eq:berry}). In terms of $\alpha$,
\begin{equation}
\gamma_b(z)=  \frac{1}{2} i \log{\left[  \frac{\cos{(\alpha(z))}}{\cos{(\alpha(0))}} \right]}  .  
\end{equation}
Substituting $\cos{(\alpha(z))}=\sqrt{1-\gamma^2/\kappa^2(z)}$, we find
\begin{equation}\label{eq:berryfinal}
\gamma_b(z)=\frac{1}{2} i \log{\bigg[\frac{\sqrt{1-\gamma^2/\kappa^2(z)}}{\sqrt{1-\gamma^2/\kappa^2(0)}}\bigg]},
\end{equation}
which is clearly a purely imaginary function below the $\mathcal{PT}$-symmetry breaking point $k(z)=\pm\gamma$, which must be upheld for all $z$. As a consequence of this, eigenstates no longer gain a phase factor on evolution, but are multiplied by a real, periodic function: $e^{i\gamma_b(z)}$.

One is able to see evidence of this function's influence when examining the spectrum of our system's instantaneous eigenstates after allowing them to evolve along a waveguide of length $L$, in the form of an instability. Ultimately, the appearance of new spectral sidebands in Fourier space will be due to the deformations of the `instantaneous wavenumber' ($\equiv \gamma_b'(z)$), similar to what happens in nonlinear systems, although Eqs. (\ref{eq:coup}) are of course fully linear; the non-Hermitian Hamiltonian Eq. (\ref{eq:ham}) is non-conservative and could give rise to instabilities, and in this case the instability comes from the fact that $\gamma_b'(z)$ is imaginary. 

In the results which follow, we choose, as a representative example, the following periodic evolution for the coupling coefficient:
\begin{equation}
\kappa(z)=1-a+a\cos{(k_0 z)},
\end{equation}
where $a$ and $k_0$ are real, positive parameters.
It is easy to see that the minimum value of this function is $1-2a$, which allows us to find a minimum threshold at which $\mathcal{PT}$-symmetry will break in terms of the parameter $a$:
\begin{equation}\label{break}
a=\frac{1-\gamma}{2}.
\end{equation}
Of course, this is based on the assumption that the adiabatic condition is upheld. If, as can be expected for a non-adiabatically evolving system, the eigenstates do not remain in their instantaneous forms, then of course we cannot as easily predict the point in parameter space where they will coalesce.

\subsection{Non-adiabatic evolution}
For a periodic coupling $k(z)$ which changes too rapidly in $z$, we can expect the requirement (\ref{eq:adiabatic}) will not be fulfilled. In this case, the eigenvalues of the system will drift from those corresponding to the instantaneous eigenstates, which will consequently change the expected $\mathcal{PT}$-symmetry breaking point. For our system, the condition for adiabatic evolution takes the form,
\begin{equation}
\bigg|\frac{1}{2}\frac{\gamma \kappa'(z)}{\kappa^2(z)-\gamma^2}\bigg|\ll|2\sqrt{\kappa^2(z)-\gamma^2}|,
\end{equation}
where $k'(z)$ is the $z$-derivative of the coupling constant $k$.
This can be conveniently rearranged to place a constraint on $k'(z)$, allowing us to determine whether the system is likely to be evolving adiabatically based on its rate of change:
\begin{equation}\label{eq:ineq}
\bigg|\frac{d}{dz}\kappa(z)\bigg|\ll\bigg|\frac{4}{\gamma} (\kappa^2(z)-\gamma^2)^{3/2}\bigg|.
\end{equation}

Even for a slowly varying coupling $\kappa(z)$, characterised by a small value of $k_0$, as our system is non-Hermitian we can expect it to display non-adiabatic behaviour as we approach the exceptional point \cite{Uzdin}. Hence, changing $k_0$ will only make significant difference in whether the system appears to be evolving adiabatically for small values of $a$. If $a$ is too close to the value given by Equation (\ref{break}), the value of $k_0$ is irrelevant. 

It is worth mentioning at this stage that a non-adiabatic generalisation of the Berry phase was introduced for cyclic Hamiltonians by Aharonov and Anandan \cite{AandA}, and extended to dissipative systems by Garrison and Wright \cite{GandW}. In the vicinity of degeneracies related to exceptional or diabolical points, where the condition (\ref{eq:adiabatic}) is violated, one can instead use the definition given in \cite{Nesterov}. However, in this work, we will continue to use the definition given by Berry for non-Hermitian systems, seen in Equation \ref{eq:berry}, as it does not require the system to be bi-orthonormal (which is violated at exceptional points), and should still be observable, even if there are additional effects due to non-adiabatic behaviour present in the results.

\section{Instability from the Geometric Phase}
\begin{figure}[h]
\centering
\includegraphics[width=0.5\textwidth]{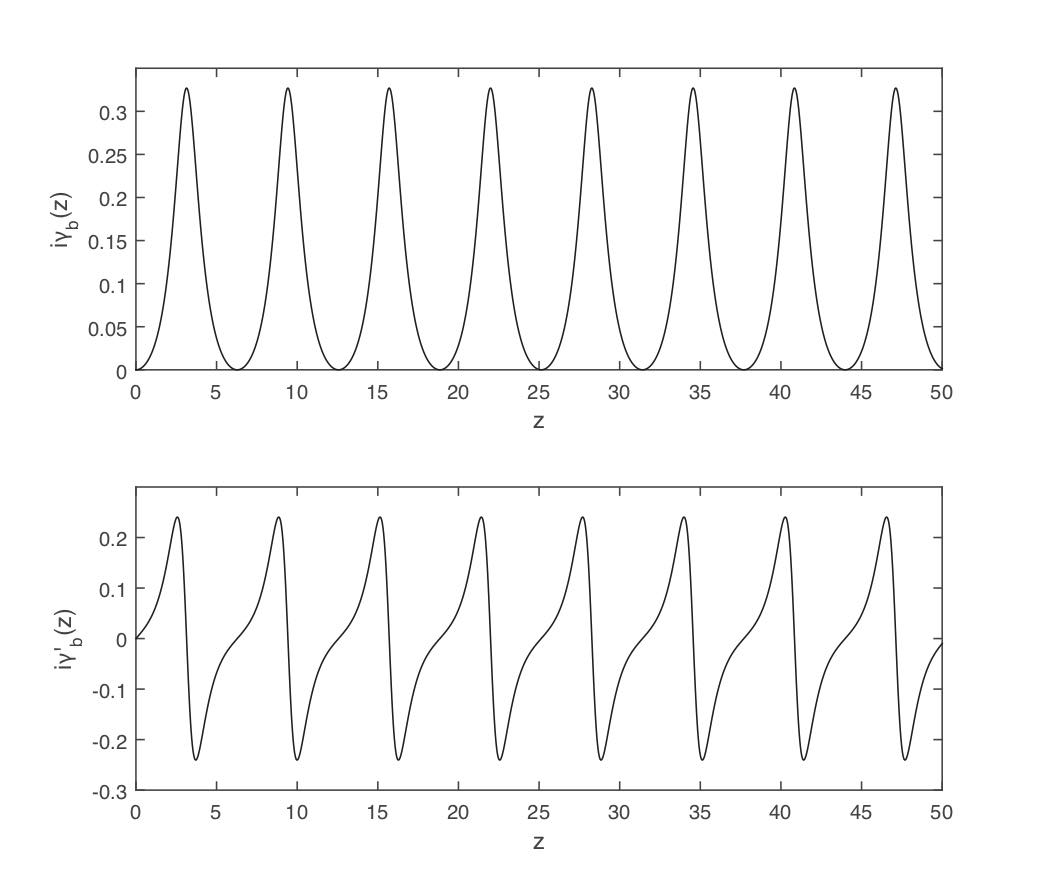}
\caption{(a) Plot of $i\gamma_b(z)$, as given by Equation (\ref{eq:berryfinal}), and (b) its $z$ derivative. In both cases, $k_0=1$, $a=0.22$, and $\gamma=0.5$. For increasing values of $a$, the amplitude of $i\gamma_b(z)$ and $i\gamma_b'(z)$ will increase, and $i\gamma_b'(z)$ will become increasingly deformed. Both functions become singular at $a=0.25$, which is the exceptional point of the system when $\gamma=0.5$, as given by Equation (\ref{break}). }
\label{fig1}
\end{figure}

We now present results of numerical simulations of Eqs. (\ref{eq:coup}), solved with a 4th-order Runge-Kutta algorithm, which demonstrate that the appearance of additional sidebands in the spectrum of states evolved along the coupled waveguides is due to the imaginary geometric function given by Equation (\ref{eq:berryfinal}). The geometric function $i \gamma_b(z)$, as given by Equation (\ref{eq:berryfinal}), and its derivative $i \gamma_b'(z)$ are shown in Figure \ref{fig1} (a) and (b) respectively for $\gamma=0.5$, $k_0=1$, and $a=0.2$. The deformed appearance of $i \gamma_b(z)$ increases with increasing values of the parameter $a$, as does the amplitude of both functions, and both functions will become singular at the exceptional point of the system, $a=0.25$, which can be found using Equation (\ref{break}). 

\begin{figure}[h]
\centering
\includegraphics[width=8cm]{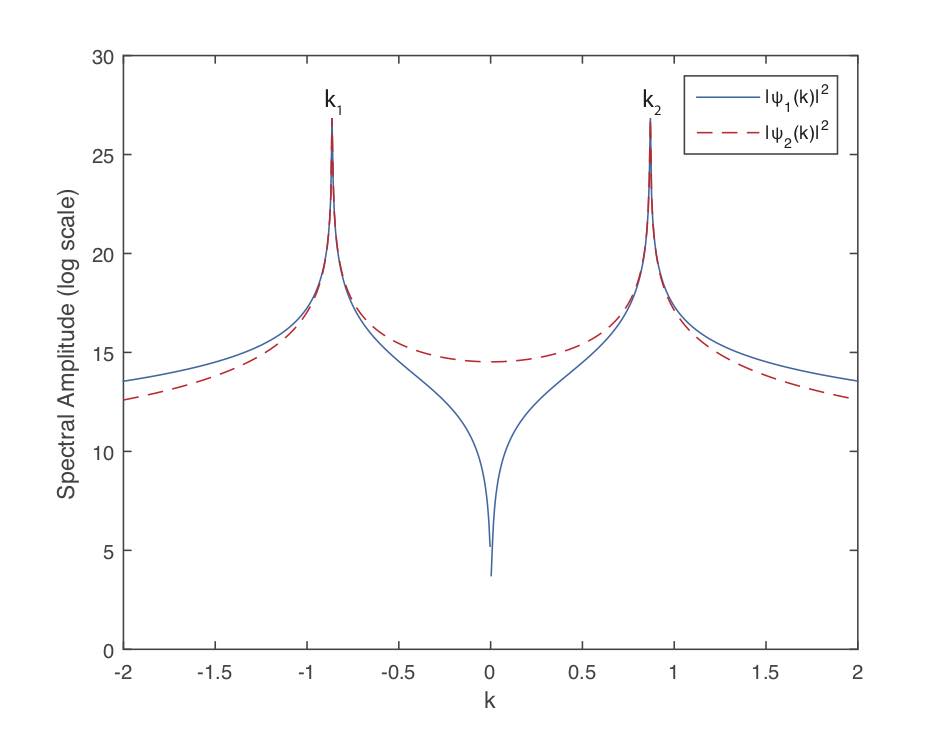}
\caption{Plot of the logarithmic spectrum of the waveguide modes $\psi_1$ and $\psi_2$, when both eigenstates are excited in a superposition $\ket{\psi_1}+\ket{\psi_2}$, and $a=0$. The right hand peak corresponds to $\ket{\psi_1}$ and the left to $\ket{\psi_2}$, and we can label them $k_1$ and $k_2$ respectively. }
\label{fig2}
\end{figure}

In our simulation, if a superposition of both eigenstates $\ket{\psi_1}+\ket{\psi_2}$ is initially excited and then propagated along the coupled waveguides described by Equations (\ref{eq:coup}a) and (\ref{eq:coup}b) for a waveguide distance $L=1000$, then for $a=0$ one will see two peaks in Fourier space, as seen in Figure \ref{fig2}. Both eigenstates remain in their instantaneous forms, as should be expected for a constant coupling constant $\kappa$. The positive peak corresponds to $\ket{\psi_1}$ and the negative to $\ket{\psi_2}$, and we shall label them $k_1$ and $k_2$ respectively. In $z$-space, if only $\ket{\psi_1}$ or $\ket{\psi_2}$ is excited, then there will be no oscillation. If we excite the superposition, then there will be a visible oscillation in $z$-space due to the envelope frequency created by the presence of $k_1$ and $k_2$. Consequently, the oscillation period in $z$ will be given by $2\pi/(k_1-k_2)$.

\begin{figure}[h]
\centering
\includegraphics[width=9cm]{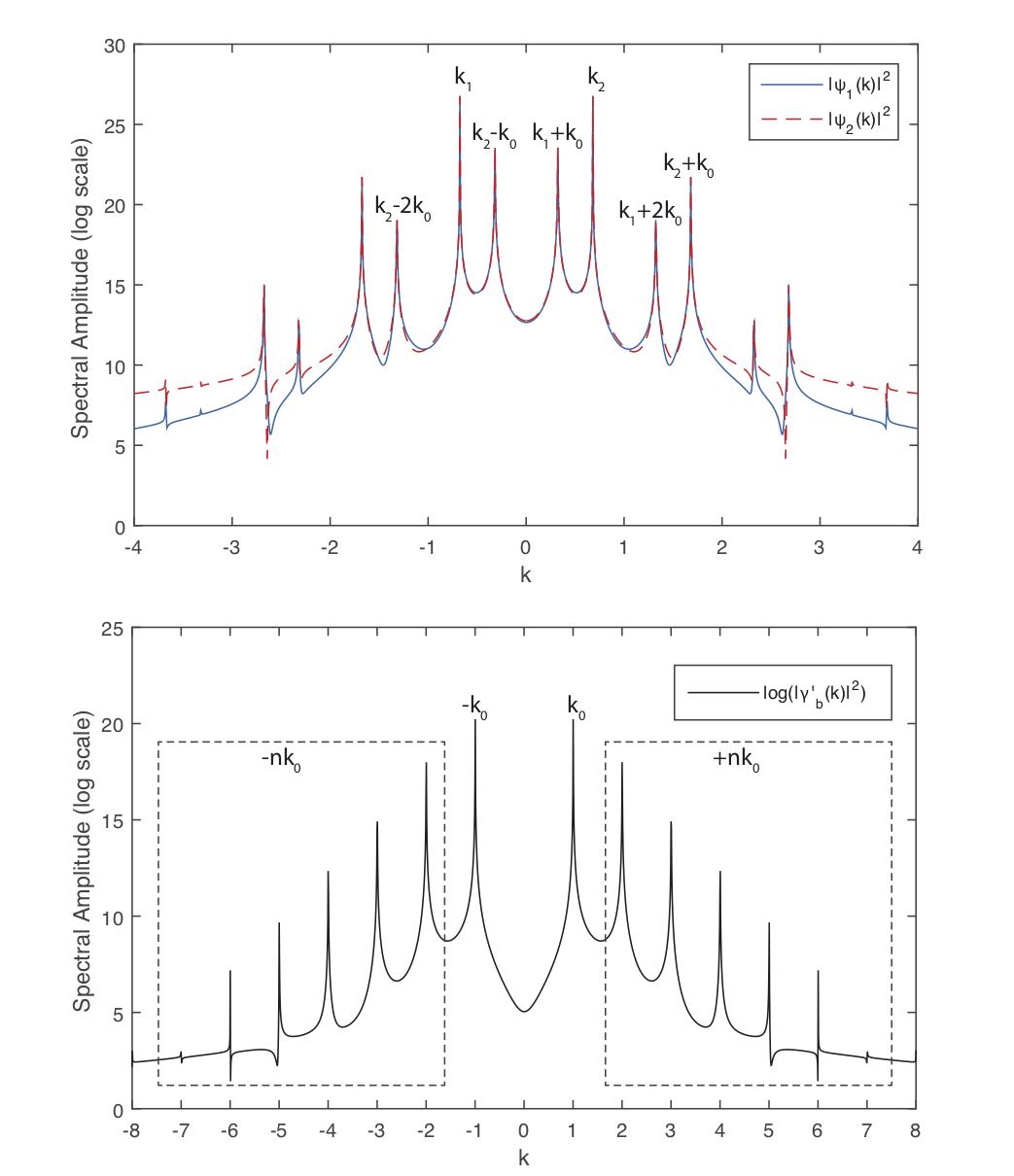}
\caption{(a) The spectrum seen when the initial state of the system is input as  $\ket{\psi_1}+\ket{\psi_2}$ for $a=0.15$, $k_0=$ and $\gamma=0.5$. The two highest peaks correspond to $k_1$ and $k_2$, although the separation between the two has decreased in comparison to Figure \ref{fig2}. (b) The spectrum of the function $i\gamma_b'(z)$, as seen in Figure \ref{fig1}. The peaks are found at $\pm \,n\, k_0$, where $n$ is a non-zero integer. Despite the fact $a=0.15$ is not very close to the exceptional point of the system, the instability peaks have a very strong presence. They can be found at $k_1 \pm n\, k_0$ and $k_2 \pm n \,k_0$, suggesting the geometric multiplier is the source of the instability peaks, as predicted.   }
\label{fig3}
\end{figure}

Figure \ref{fig3} (a) shows the logarithmic spectrum seen after propagation along the coupled waveguides of $L=1000$ when exciting a superposition of both eigenstates for $a=0.15$, $k_0=1$, and $\gamma=0.5$. The two highest peaks correspond to $k_1$ and $k_2$, and as predicted in the previous sections we also see instability peaks. In Figure \ref{fig3} (b) we can clearly see that the peaks correspond to the points $\pm \,n \, k_0$ in Fourier space, where $n\in\mathbb{N}$. As a result, the instability peaks in Figure \ref{fig3} (a) correspond to $k$ values of $k_1 \pm n\,k_0$ and $k_2 \pm n\, k_0$. In comparison with Fig. \ref{fig2}, the separation between $k_1$ and $k_2$ has decreased. This is in line with the expectation that the two instantaneous eigenstates of the system will approach each other as the exceptional point is approached.  

\begin{figure}[h]
\centering
\includegraphics[width=0.5\textwidth]{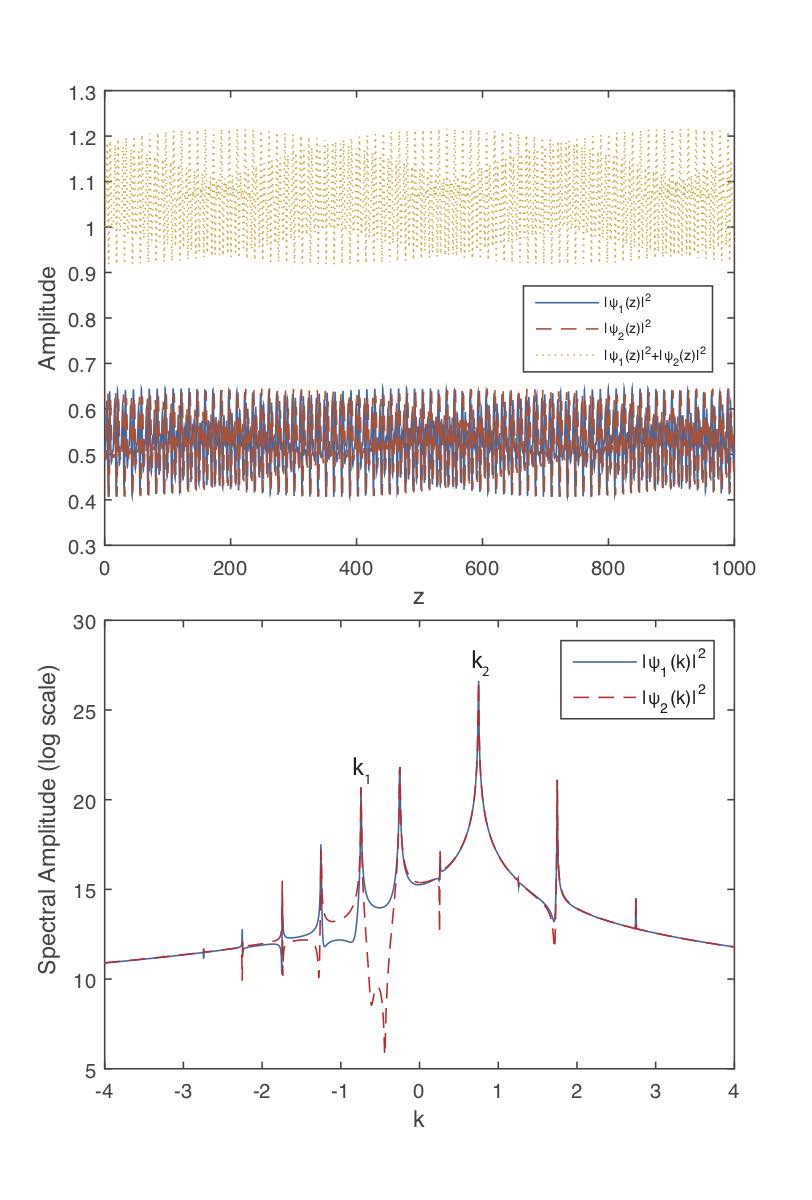}
\caption{(a) Evolution in $z$ when the initial state is the instantaneous eigenstate $\ket{\psi_1(0)}$ for $a=0.1$, $k_0=1$, and $\gamma=0.5$ of $|\psi_1|^2$ (blue solid line), $|\psi_2|^2$ (red dashed line), and their sum, equivalent to the norm squared of the evolved eigenstate $|\braket{\psi_1(z)}{\psi_1(z)}|^2$ (yellow dotted line). The oscillations in $z$-space are periodic, but also deformed, displaying evidence of the influence of multiple frequencies and the presence of the geometric multiplier $e^{i \gamma_b(z)}$. (b) The logarithmic spectrum taken at $L=1000$ corresponding to the data shown in (a). The spectrum no longer displays the symmetry seen before, due to the fact only a single eigenstate is excited. Nevertheless, a smaller peak corresponding to $k_2$ is present suggesting a departure from adiabatic evolution, as well as the majority of the expected instability peaks.   }
\label{fig4}
\end{figure}

In Figure \ref{fig4}, the $z$-space oscillations and corresponding spectrum at $L=1000$ is seen when only $\ket{\psi_1}$ is excited for $a=0.1$. As $a$ is fairly small and reasonably far from the exceptional point ($a=0.25$), the influence of the instability peaks is not as strong as that seen in Figure \ref{fig3}. Furthermore, as only $\ket{\psi_1}$ is initially excited the spectrum is asymmetric. Nevertheless, the presence of the $k_2$ peak is clear, although it is not as strong as the peak $k_1$, and could be easily mistaken for an instability peak. Its appearance suggests the system is not evolving adiabatically, which is to be expected when examining the inequality (\ref{eq:ineq}): as $a$ grows the $\kappa(z)$ derivative will gradually  become comparable in magnitude with the inequality's right hand side. Examining the $z$-evolution of $|\psi_1|^2$ and $|\psi_2|^2$ in Figure \ref{fig4} (a) will again reveal the appearance of oscillations due to the simultaneous presence of $k_1$ and $k_2$. These appear similar to the oscillations seen in \cite{Rabi}, known as Rabi oscillations, where gain and loss, rather than the coupling, are driven. It should be noted that in unlike those seen in \cite{Rabi}, the oscillations of $|\psi_1|^2$ and $|\psi_2|^2$ are not perfectly out of phase, and also appear to show some deformation, suggesting they are being influenced by $i \gamma_b'(z)$, as seen in Figure \ref{fig1} (b).

\begin{figure}[h]
\centering
\includegraphics[width=0.5\textwidth]{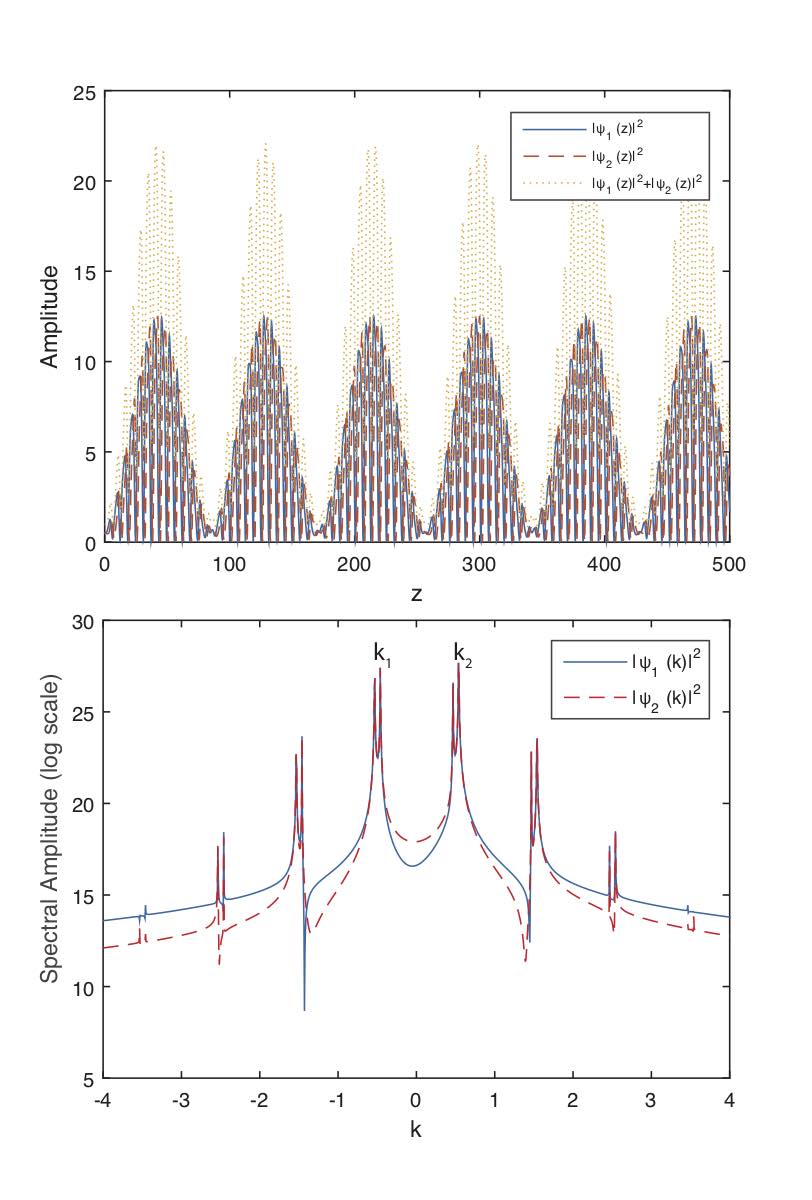}
\caption{(a) Evolution in $z$ when the initial state is the instantaneous eigenstate $\ket{\psi_1(0)}$ for $a=0.22$, $k_0=1$, and $\gamma=0.5$ of $|\psi_1|^2$ (blue solid line), $|\psi_2|^2$ (red dashed line), and their sum, equivalent to the norm squared of the evolved eigenstate $|\braket{\psi_1(z)}{\psi_1(z)}|^2$ (yellow dotted line). The oscillations in $z$-space are still present, and a clear beating enveloping the smaller oscillations has appeared. (b) The spectrum taken at $L=1000$ corresponding to the data shown in (a). The logarithmic spectrum is now symmetric, despite no initial excitation of $\ket{\psi_2}$, which implies a complete departure from adiabatic behaviour. The separation between all peaks has decreased, and the smallest separation between peaks can be directly linked to the beating seen in $z$-space, as explained in the main text. }
\label{fig5}
\end{figure}
\begin{figure}[h]
\centering
\includegraphics[width=0.5\textwidth]{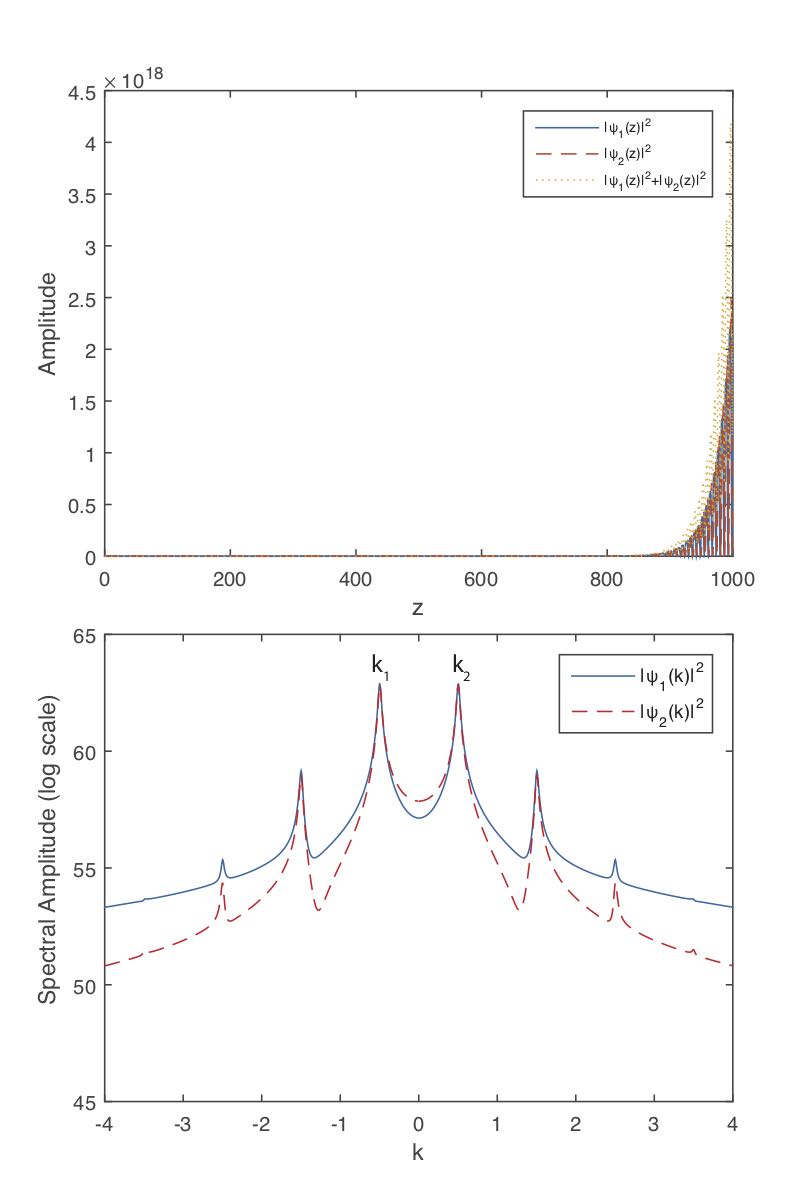}
\caption{(a) Evolution in $z$ when the initial state is the instantaneous eigenstate $\ket{\psi_1(0)}$ for $a=0.225$, $k_0=1$, and $\gamma=0.5$ of $|\psi_1|^2$ (blue solid line), $|\psi_2|^2$ (red dashed line), and their sum, equivalent to the norm squared of the evolved eigenstate $|\braket{\psi_1(z)}{\psi_1(z)}|^2$ (yellow dotted line). There is clear exponential gain, suggesting $\mathcal{PT}$-symmetry has been broken and the energies of the system have become imaginary. (b) The logarithmic spectrum taken at $L=1000$ corresponding to the data shown in (a). The separation between the two central peaks is now equal to $k_0$, and the previously approaching peaks have now overlapped.}
\label{fig6}
\end{figure}
As $a$ is increased, a strong beating in $z$-space will appear, and despite exciting only one eigenstate initially, both $k_1$ and $k_2$ will appear in the spectrum with equal strength, displaying a complete breakdown of adiabatic evolution. This is shown in Figure \ref{fig5} for $a=0.22$, $k_0=1$, and $\gamma=0.5$, where again only $\ket{\psi_1}$ has been initially excited.  The envelope of the smaller oscillations observed is characteristic of the $\mathcal{PT}$-symmetry breaking point being approached, and the amplitude of this envelope will grow for increasing values of the parameter $a$.  Furthermore, one should note that due to the decrease in the separation between $k_1$ and $k_2$, the instability peaks have moved much closer to one another, and are approaching the point of overlap. The separation between the maxima of the envelope wave observed in $z$-space can be shown to be approximately equal to $2\pi$ divided by the smallest separation of peaks in the spectrum, e.g. the separation between two instability peaks in Fourier space. 

In the simulation, the $\mathcal{PT}$-symmetry of the system breaks at approximately $a=0.223$ for $\gamma=0.5$ and $k_0=1$. This is characterised by the onset of exponential gain in $z$-space, suggesting the exceptional point of the system has been crossed and the energies of $\ket{\psi_1}$ and $\ket{\psi_2}$ have become imaginary. The evolution of the system after this point is crossed can be seen in Figure \ref{fig6} (a), with the corresponding spectrum shown in Figure \ref{fig6} (b), for $a=0.225$, $\gamma=0.5$, and $k_0=1$. The early onset of the broken $\mathcal{PT}$-symmetry phase (compared to the theoretically predicted break point of $a=0.25$) should not be surprising given the clearly non-adiabatic behaviour of the system; the drift of $\ket{\psi_1}$ and $\ket{\psi_2}$ from their instantaneous forms makes the precise point at which they will coalesce very difficult to predict with the estimates we give in previous sections. In Figure \ref{fig6} (b) we can see that the instability peaks have now overlapped, and the separation between $k_1$ and $k_2$ has reduced, such that it's equal to the parameter $k_0$. This is not a coincidence - changing $k_0$ will determine the peak separation at the $\mathcal{PT}$-symmetry breaking point observed in the simulation.

In the results so far we have only seen the behaviour of our coupled waveguide system for $k_0=1$, which can be freely chosen. It may seem that by reducing this value, one can better ensure that the inequality (\ref{eq:ineq}) is upheld, and likewise by increasing it ensure the opposite, enabling one to find a case where the system displays $\mathcal{PT}$-symmetry breaking at the expected point of $a=0.25$, in a similar vein to what is seen in \cite{Jaouadi}. However, in practice this does not work: close to the exceptional point, no matter how slow the variation of the coupling in $z$ is, the system will not behave adiabatically, making it very difficult to predict accurately where $\mathcal{PT}$-symmetry will break in the simulation. This is in line with what is already known about non-Hermitian systems, as reported in \cite{Uzdin}, where it is shown that the gain inherently present in non-Hermitian systems will magnify usually neglected adiabatic effects.

\section{Conclusions}
It is shown that the eigenstates of a non-Hermitian, $\mathcal{PT}$-symmetric coupled waveguide system with a periodically varying coupling will gain a purely imaginary geometric multiplier on evolution, which will in turn induce an instability in the system, visible in the spectrum of the evolved eigenstates. The magnitude of the instability will grow as the system approaches its $\mathcal{PT}$-symmetry breaking point, which is observed breaking for a smaller value of the parameter $a$ than theoretically predicted. This is expected, as the conditions required for adiabatic evolution in non-Hermitian systems is violated in the vicinity of the exceptional point. Beyond the $\mathcal{PT}$-symmetry breaking point, exponential gain will be observed in the coupled waveguides. This new mechanism of instability has interesting implications for the optics community. Learning how to control the output of a waveguide system by modulating the distance between the waveguides (and thus modulating the coupling) could lead to new switching or routing devices based on the complex Berry phase \cite{lupu}. Furthermore, it is possible that such an instability will also be visible in other branches of physics in which non-Hermitian two-level systems can be found, such as photon fluids, where complex Berry phases are also known to arise \cite{Nic}.

\section{Acknowledgements}
The authors would like to acknowledge funding from the EPSRC Centre of Doctoral Training for Condensed Matter Physics (CM-CDT).

\end{document}